\documentclass[a4paper,12pt]{article}

\usepackage{jheppub} 

\usepackage{amssymb}
\usepackage{amsmath}
\usepackage{amsthm}
\usepackage{mathrsfs}
\usepackage[numbers]{natbib}
\usepackage{mathtools}
\usepackage{graphicx}
\usepackage{float}
\usepackage{subfigure}
\usepackage{booktabs}
\usepackage{multirow}
\usepackage[colorlinks=true]{hyperref}
\usepackage{physics}
\usepackage{slashed}
\usepackage{diagbox}

\makeatletter
\gdef\@fpheader{}
\makeatother

\title{Electric Dipole Moments from Axion-Like Particle Dark Matter Background}

\author[a,b]{Jason L. Evans}
\author[a,b]{Nan Li}
\author[a,b]{Junyuan Lyu}
\affiliation[a]{Tsung-Dao Lee Institute, Shanghai Jiao Tong University, Shanghai 201210, China}
\affiliation[b]{School of Physics and Astronomy, Shanghai Jiao Tong University, Shanghai 200240, China}

\emailAdd{jlevans@sjtu.edu.cn}
\emailAdd{nanli2000cn@sjtu.edu.cn}
\emailAdd{lvjunyuan@sjtu.edu.cn}

\abstract{New one- and two-loop contributions to the lepton's and nucleon's EDM, which are induced by an axion-like particle dark matter background, are examined. These contributions include effects from CP-violating ALP interactions with photons, leptons and nucleons. The contribution to the EDM is so larger it leads to new constraints on the CP violating couplings of axion-like particles, if the axion-like particle mass is smaller than $10^{-11} $ eV.}

\begin{document}
\maketitle
\flushbottom

\section{Introduction}
The nature of dark matter is still very much in dispute at the most fundamental level.  One particularly promising dark matter candidate is the quantum chromodynamics (QCD) axion. This candidate is a particularly interesting possibility, since it can also solve the the strong CP problem\cite{axion 1,axion 2,axion 3,axion 4,axion 5}. However, standard QCD axion models predict a strict relation between the axion mass and its decay constant, if it is to be dark matter. A more generic string inspired version of the QCD axion, usually refereed to as an axion-like particle (ALP) \cite{axion 6,axion 7}, has a more generic parameter space for which it is a viable DM candidate\cite{axion 8,axion 9,axion 10,axion 11}. 

In this work, we focus on ALPs with masses below an eV. For models with ALP masses well below an eV, the large De Broglie wavelength of the dark matter helps solve the core-cusp problem\cite{UDM astro 1,UDM astro 2,UDM astro 3,UDM astro 4,UDM astro 5}. The long de Broglie wavelength of these particles sets a lower bound on the size of the dark matter density fluctuations in the early universe forcing a cored dark matter profile. 

It was recently pointed out \cite{Background UDM 1} that the background effects of ultralight dark matter, which has a huge occupation number, can have important consequences for the predicted value of standard model observables. This work showed that background enhancements to the loop contribution of the electron $g-2$ were so large that experimental measurements excluded a dark photon kinetic mixing parameter larger than $\chi> 7.1\times 10^{-11}(m_{\gamma}'/10^{-14} \,\mathrm{eV})$. This gives the strongest constraint on the gauge mixing parameter when  $m_{\gamma}' \lesssim  10^{-14}\mathrm{eV}$ , if the dark photon is dark matter. As pointed out in \cite{Background UDM 1}, these background effects must be taken into account for any process involving ultralight dark matter. Here, we will show that this background effect is also important for ALPs and there CP-violating interactions. The enhancement factor arising from the background is generally of order  $\rho_{\phi}/(m_{\phi}^{2}m_{i}^{2})$ and can be larger than one for first generation particles as long as $m_\phi\lesssim 10^{-10}$ eV, where $m_\phi$ is the ultralight dark matter mass and $m_{i} $ is the mass of the fermion interacting with the dark matter background. 

Because this enhancement to the loop processes coming from the background is fairly generic and robust, it should also effect contributions to things like the electric dipole moment (EDM). This effect is examined here.  

Since we are examining low energy contributions, we use effective operator interactions of the ALPs with Standard Model fermions and gauge bosons for scales below the GeV scale \cite{d=5 Lagrangian}, which are determined by chiral perturbation theory ($\chi $pt). The resulting effective interactions of the photon, leptons and nucleons in $\chi $pt can then be constrained by experiments that measure EDMs.\cite{EDM Experiment 1,EDM Experiment 2,EDM Experiment 3}

In this work, we will consider EDM contributions involving the background up to the two-loop order. These diagrams will give the leading order contribution to various EDMs for different CP-violating couplings, such as ALP-photon-photon and ALP-fermion-fermion.  We will consider both lepton EDM and nucleon EDM. 

This article is organized as follows: In section~\ref{section 2} we will discuss generic aspects of background corrections to the EDMs. In section~\ref{section 3}, we will calculate background effects on the lepton EDM. Then, in section~\ref{section 4}, we will focus on nucleon EDMs. Next, section~\ref{section 5} uses recent experiment limits on $d_{e},d_{n},d_{p} $ to find new constraints on the ALP CP-violating interactions with the photon, leptons and nucleons. Our conclusions are presented in section~\ref{section 6}. 


\section{Electric Dipole Moments in a Background}
\label{section 2}
The dark matter density near earth has been measured with a value of approximately $0.3~{\rm GeV/cm^3}$. Since the number density scales as the dark matter density divided by mass, the number of particles need to realize this energy density increase as the dark matter candidate mass decrease. For ultralight dark matter, this number is huge. To approximately convert the number density to the occupation number, the number density is divided by the momentum space volume. Since dark matter is non-relativistic, this leads to an incredibly large occupation number. With the occupation number density $n(k) $ is so large,  the ground state can no longer be described by the a simple vacuum state, $\ket{0} $. Instead the ground state must be some $n(k)$ particle state, $\ket{n(k)} $.  This non-trivial background state then modifies the propagator of the ALP \cite{Background UDM 2} to 
\begin{align}
    \tilde{\Delta}_{\phi}(k^{2})=\mathcal{F}\{\mel{n(k)}{T\phi(x)\phi(0)}{n(k)}\}=\frac{\mathrm{i}}{k^{2}-m_{\phi}^{2}+\mathrm{i}\epsilon}-2\pi\,n(k)\,\delta\big(k^{2}-m_{\phi}^{2}\big),
\end{align}
where $\mathcal{F} $ signifies Fourier transformation. Because the dark matter is quite non-relativistic, the following approximate form for the occupation number density $n(k)$ can be used in most calculations \cite{Background UDM 3},
\begin{align}
    n(k)=\frac{\rho_{\phi}}{m_{\phi}}(2\pi)^{3}\delta^{3}(\vb{k}),
    \label{occupation number density}
\end{align}
where $\rho_{\phi} $ is the mass density of ALP in the universe and $m_\phi$ is the ALP mass\footnote{There are corrections proportional to the dark matter velocity but are subdominant for the cases we consider.}. Here, we will only consider background dependent contributions, i.e. the part of the propagator proportional to the occupation number density. The contribution arising from the free theory is well known and reviewed in \cite{CPV Axion review}.

As is well known, the effective Lagrangian for an EDM can be written as
\begin{align}
  \mathcal{L}_{\mathrm{EDM}}=-\frac{\mathrm{i}}{2}d\,\bar{\psi}\sigma^{\mu\nu}\gamma_{5}\psi F_{\mu\nu}
  =-d\,\bar{\psi}\gamma^{\mu\nu}\gamma_{5}\psi(\partial_{\nu}A_{\mu}),
\end{align}
where $\sigma^{\mu\nu}=\mathrm{i}\gamma^{\mu\nu}=\mathrm{i}[\gamma^{\mu},\gamma^{\nu}]/2 $ and $d $ is the electric dipole moment. The loop-level matrix element of the electromagnetic current which leads to an EDM contribution takes the following form,
\begin{align}
    \mel**{\vb{p}_{2},s_{2}}{J^{\mu}(x)}{\vb{p}_{1},s_{1}}_{\mathrm{EDM}}=&-d\,\bar{u}_{s_{2}}(p_{2})q_{\nu}\gamma^{\mu\nu}\gamma_{5}u_{s_{1}}(p_{1})~,\notag\\
    =&-2d\,\bar{u}_{s_{2}}(p_{2})\bar{p}^{\mu}\gamma_{5}u_{s_{1}}(p_{1})~.
    \label{standard EDM matrix element of J}
\end{align}
where $\bar{p}=(p_{1}+p_{2})/2 $, $q=p_2-p_1$ and a variant of Gordon's Decomposition has been use to the get the second line. If a background is present, Lorentz invariance is broken and the matrix elements of the electromagnetic current giving an EDM can have a non-standard form \eqref{standard EDM matrix element of J}.

Since the background breaks the Lorentz symmetry, we must be careful about which reference frame the EDM is calculated in. The EDM is generally measured in a uniform electric and magnetic field \cite{EDM Experiment 1,EDM Experiment 2,EDM Experiment 3}, thus, this is the reference frame the EDM is calculated in. However, the background does not break gauge invariance \cite{Background UDM 1}. Thus, we are free to choose the gauge $A^{0}\neq 0, A^{i}\simeq 0 $\footnote{The electric and magnetic field in experiment \cite{EDM Experiment 1} are respectively $78 \;\mathrm{GV}/\mathrm{cm}$ and $<100 \;\mathrm{mGauss}$, so $A^{i}/A^{0}< 10^{-9}$.}. The gauge invariance of each contribution to the EDM will be discussed in Appendix \ref{appendix a} and proved on more general grounds in Appendix \ref{appendix b}. Matrix elements of the interaction Hamiltonian related to the EDM, which are proportional to  $\vb{S}\cdot \vb{E}$, stem from the matrix elements of $\int\dd[3]x\;\langle J^\mu\rangle A_\mu$, where $A_{\mu} $ is an external field. Therefore, gauge invariance allows to consider only the 0-th component of this matrix elements.

Even after considering the effect of the background, the electric charge is not renormalized and the standard quantum electrodynamic interaction can be used.  Furthermore,  the fermion fields can also be taken to be the standard Dirac fields\cite{Background UDM 1,Background UDM 2,Background UDM 3}. Specifically, the  spinors $u_{s}(\vb{p}), v_{s}(\vb{p}) $ and the electron-photon vertex, $-\mathrm{i}e\gamma^{\mu} $, are the conventional values \footnote{Corrections to the fermion propagators from background dark matter do not induce an EDM and these contributions are ignored, since they are subleading.}.

\section{Lepton EDM}
\label{section 3}

Before we calculate any electric dipole moments, we will define our theory. The most general $SU(3)_{\mathrm{c}}\times U(1)_{\mathrm{em}} $ invariant $d=5 $ effective Lagrangian below the electroweak scale containing CP-violating ALP interactions with the photon, gluons and SM fermions, is \cite{d=5 Lagrangian,CPV Axion review,d=5 Lagrangian 2}

\begin{align}
    \mathcal{L}_{\phi}\supset &e^{2}\frac{\tilde{C}_{\gamma}}{\Lambda}\phi F\tilde{F}+\mathrm{i}\frac{v}{\Lambda}y_{\mathrm{P}}^{ij}\phi \bar{f}_{i}\gamma_{5}f_{j}+g_{s}^{2}\frac{\tilde{C}_{g}}{\Lambda}\phi G\tilde{G}\notag\\
    &+e^{2}\frac{C_{\gamma}}{\Lambda}\phi FF+\frac{v}{\Lambda}y_{\mathrm{S}}^{ij}\phi \bar{f}_{i}f_{j}+g_{s}^{2}\frac{C_{g}}{\Lambda}\phi GG,
    \label{general Lagrangian}
\end{align}
where $f\in\{e,u,d\} $ denotes SM fermions in the mass basis, $i,j $ are flavor indices, and $y_{\mathrm{S}}, y_{\mathrm{P}} $ are hermitian matrices for the scalar and pseudoscalar type interactions. $F $ and $G $ are the QED and QCD field strength tensors respectively, while $\tilde{F} $ and $\tilde{G} $ are their duals. If non-perturbative effects from the boundary terms are neglected, the first line of \eqref{general Lagrangian} will preserve the $\phi$ shift symmetry. The second line of \eqref{general Lagrangian}, however, will break the shift symmetry. Thus, the results of our EDM calculation will not preserve the axion-like particle shift symmetry.

CP-violating phenomena are described by Jarlskog invariants\cite{Jarlskog invariant}, i.e. reparametrization invariant combination of parameters. The relevant Jarlskog invariance for our considerations are 
\begin{align}
    y_{\mathrm{S}}^{ij}y_{\mathrm{P}}^{ji},\qquad y_{\mathrm{S}}\tilde{C}_{\gamma},\qquad y_{\mathrm{P}}C_{\gamma},\qquad y_{\mathrm{S}}^{ii}y_{\mathrm{P}}^{jj},\qquad y_{\mathrm{S}}^{jj}y_{\mathrm{P}}^{ii},\qquad C_\gamma \tilde{C}_\gamma.
\end{align}

If the energy scale associated with the dark matter particle is lower than about 1 GeV, non-perturbative methods must be invoked for strongly coupled particles and chiral perturbation theory is employed

\begin{align}
    \mathcal{L}_{\phi}^{\chi\mathrm{pt}}\supset e^{2}\frac{\tilde{c}_{\gamma}}{\Lambda}\phi F\tilde{F}+\mathrm{i}\frac{v}{\Lambda}y_{\mathrm{P}}^{ij}\phi \bar{\ell}_{i}\gamma_{5}\ell_{j}
    +e^{2}\frac{c_{\gamma}}{\Lambda}\phi FF+\frac{v}{\Lambda}y_{\mathrm{S}}^{ij}\phi \bar{\ell}_{i}\ell_{j},
    \label{Lagrangian}
\end{align}
where $i,j $ are flavor indices of lepton. Even though chiral perturbation theory is unimportant for the lepton sector and the relevant interactions can be read directly from Eq. \eqref{general Lagrangian}, $c_\gamma$ and $\tilde{c}_\gamma$ are still different from $C_\gamma$ and $\tilde{C}_\gamma$ at low energy scales, see eq.(50) of \cite{CPV Axion review}. To help the redear remeber this, we use $c_\gamma$ and $\tilde{c}_\gamma$ instead of $C_\gamma$ and $\tilde{C}_\gamma$ in the relevant interactions.  


\subsection{Loop Calculation of EDM}
\begin{figure}[htbp]
  \centering
  \subfigure[]{\includegraphics[width=0.4\textwidth]{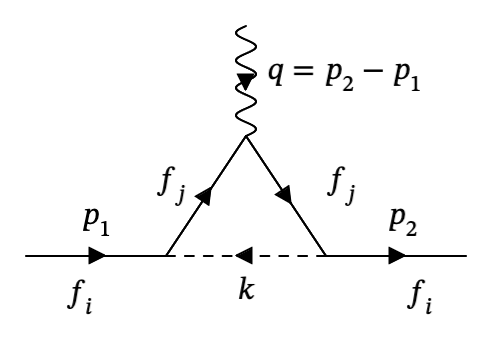}\label{1a}}
  \subfigure[]{\includegraphics[width=0.4\textwidth]{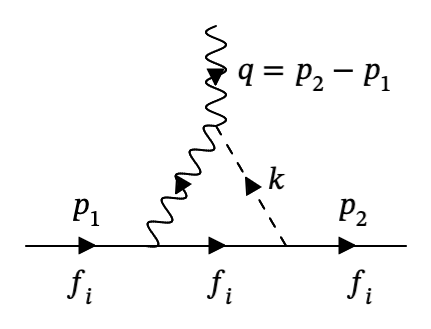}\label{1b}}\\ 
  \subfigure[]{\includegraphics[width=0.4\textwidth]{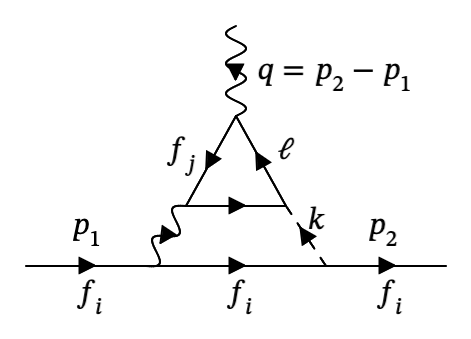}\label{1c}}
  \subfigure[]{\includegraphics[width=0.4\textwidth]{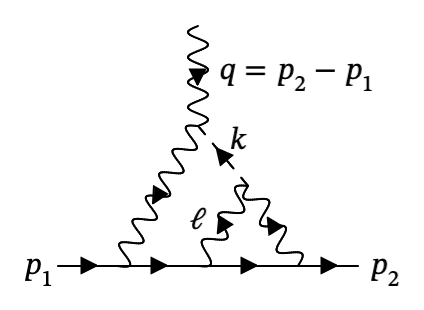}\label{1d}}
  \caption{Leading contributions to the lepton EDM}
  \label{EDM}
\end{figure}

In this section, we will calculate the leading order background dependent contribution to the lepton EDM. This contribution comes from the diagrams in Figure~\ref{EDM} with one CP-odd and one CP-even $\phi$ interaction. Each diagram in Figure~\ref{1a}-\ref{1d} are related to different combinations of coupling: (a) $y_{\mathrm{S}}^{ij}y_{\mathrm{P}}^{ji}$; (b) $y_{\mathrm{S}}\tilde{c}_{\gamma}$, $y_{\mathrm{P}}c_{\gamma}$; (c) $y_{\mathrm{S}}^{ii}y_{\mathrm{P}}^{jj}$, $y_{\mathrm{S}}^{jj}y_{\mathrm{P}}^{ii}$; (d) $c_\gamma \tilde{c}_\gamma$. We find that Figure~\ref{1c} and \ref{1d} are equivalent to Figure~\ref{1b} with the vertex replaced by the results of calculating the triangle subdiagram. Using this fact will allow us to apply the calculations of Figure~\ref{1b} to Figure~\ref{1c} and \ref{1d}.

The EDM experiments\cite{EDM Experiment 1,EDM Experiment 2,EDM Experiment 3} providing the constraints we apply have uniform electric and magnetic fields and thus have small momenta and energy in comparison to the electron mass. This will allow us to ignore $\order{q^{2}} $ effects in any calculation, i.e. take the approximation $p_{1}\simeq p_{2}\simeq \bar{p} $. In addition to this approximation, we will take $m_{\phi} $ much smaller than any mass scale in the calculation keeping only leading order terms in $m_{\phi}$.  It is this regime for which the background contribution is considerable. We also take the non-relativistic limit for the fermions, $\bar{p}^{0}\simeq m_{i} $, which again is consistent with the EDM experimental set up.

Next, we calculate the EDM contribution found in Fig. (\ref{1a}) with one CP-even and CP-odd vertex to get

\begin{align}
  \mathrm{i}M_{1,\ell}^{\mu}=&e\frac{v^{2}}{\Lambda^{2}}\sum_{j}\int{\frac{\dd[4]{k}}{(2\pi)^4}}\;\tilde{\Delta}_{\phi}(k^2)\frac{\bar{u}_{s_{2}}(p_{2})N^{\mu}_{1,\ell}u_{s_{1}}(p_{1})}{\big((\bar{p}+k)^{2}-m_{j}^{2}\big)^{2}}
  \label{lepton 1}
\end{align}
where 
\begin{align}
    N^{\mu}_{1,\ell}
    =&4\Re(y_{\mathrm{S}}^{ij}y_{\mathrm{P}}^{ji})m_{j}k^{\mu}\gamma_{5}\notag\\
  &-2\mathrm{i}\Im(y_{\mathrm{S}}^{ij}y_{\mathrm{P}}^{ji})\Big[(k^{2}+m_{i}^{2}-m_{j}^{2})\gamma^{\mu}-2k^{\mu}\slashed{k}+m_{i}[\slashed{k},\gamma^{\mu}]\Big]\gamma_{5}.
  \label{N_1l}
\end{align}
Only the first line above will lead to an EDM, see Appendix~\ref{appendix a}. As stated earlier, we take $A^{0}\neq 0, A^{i}\simeq 0 $. This means only the 0-th component of $\mathrm{i}M_{1,\ell}^{\mu}$ will contribute to the EDM at leading order. With this assumption, the EDM contribution of diagram~\ref{1a} is
\begin{align}
  \mathrm{i}M_{1,\ell\mathrm{EDM}}^{0}\simeq &2e\frac{v^{2}}{\Lambda^{2}}\sum_{j}\Re(y_{\mathrm{S}}^{ij}y_{\mathrm{P}}^{ji})\frac{m_{j}}{m_{i}}\int{\frac{\dd[4]{k}}{(2\pi)^4}}\;\frac{k^{0}\tilde{\Delta}_{\phi}(k^{2})}{\big((\bar{p}+k)^{2}-m_{j}^{2}\big)^{2}}\notag\\
  &\qquad\times\bar{u}_{s_{2}}(p_{2})q_{\nu}\gamma^{0\nu}\gamma_{5}u_{s_{1}}(p_{1}),
\end{align}
where we have further simplified using the fact that the fermions are non-relativistic, i.e. $\bar p^0 \simeq m_i$. If we then use \eqref{standard EDM matrix element of J}, we get
\begin{align}
    d_{1,\ell} \simeq -2e\frac{v^{2}}{\Lambda^{2}}y_{\mathrm{S}}^{ii}y_{\mathrm{P}}^{ii}I_{1},
\end{align}
for the EDM, where 
\begin{align}
    I_{1}=\int{\frac{\dd[4]{k}}{(2\pi)^4}}\;\tilde{\Delta}_{\phi}(k^{2})\frac{k^{0}}{(k^{2}+2k\cdot\bar{p})^{2}}~,
    \label{I_1}
\end{align}
and $\Re(y_{\mathrm{S}}^{ii}y_{\mathrm{P}}^{ii})=y_{\mathrm{S}}^{ii}y_{\mathrm{P}}^{ii}$ for hermitian $y_{\mathrm{S}},y_{\mathrm{P}}$.

With the same set of simplifications, the matrix element of Figure~{\ref{1b}} is found to be
\begin{align}
  \mathrm{i}M_{2,\ell\mathrm{EDM}}^{0}\simeq &8e^{3}\frac{v}{\Lambda^{2}}\int{\frac{\dd[4]{k}}{(2\pi)^4}}\;\tilde{\Delta}_{\phi}(k^{2})\frac{1}{(k^{2}+q^{2})(2k\cdot\bar{p}+k^{2})}\bar{u}_{s_{2}}(p_{2})\notag\\ 
  &\qquad \times\big(y_{\mathrm{P}}^{ii}c_{\gamma}(k^{0}k_{0}-k^{0}q^{2}/2m_{i})-2y_{\mathrm{S}}^{ii}\tilde{c}_{\gamma}\vb{k}^2\big)q_{\nu}\gamma^{0\nu}\gamma_{5}u_{s_{1}}(p_{1}),
\end{align}
where the full expression can be found in Appendix \ref{appendix a}. The contribution from Figure~\ref{1b}, which is proportional to $y_{\mathrm{S}}\tilde{c}_\gamma$, is suppressed by $\vb{k}^{2} $, the dark matter velocity
\footnote{The constraints on $y_{\mathrm{P}}^{ee}c_\gamma $, $c_\gamma \tilde{c}_\gamma$ and $y_{\mathrm{P}}^{ee}y_{\mathrm{S}}^{ee}$ in Table~\ref{New bounds by electron EDM} below and on $y_{\mathrm{P}}^{ee}$ in \cite{Background UDM 3}, can be combined to give a constraint on $y_{\mathrm{S}}^{ee}\tilde{c}_\gamma$.  Apriori, these constraints could lead to suppressed couplings and the contribution from $y_{\mathrm{P}}^{ee}c_\gamma$ to the EDM would become subdominant.  In this case, the momentum suppressed contribution coming from $y_{\mathrm{S}}^{ee}\tilde{c}_\gamma$ would dominant and we would have a different constraint coming from this contribution to the EDM.}.
We then get
\begin{align}
    d_{2,\ell}\simeq-8e^{3}\frac{v}{\Lambda^{2}}y_{\mathrm{P}}^{ii}c_{\gamma}I_{2},
\end{align}
where 
\begin{align}
    I_{2}=\int{\frac{\dd[4]{k}}{(2\pi)^4}}\;\tilde{\Delta}_{\phi}(k^{2})\frac{k^{0}k_{0}-k^{0}q^{2}/2m_{i}}{(k^{2}+q^{2})(2k\cdot\bar{p}+k^{2})}~.
    \label{I_2}
\end{align}

\begin{figure}[htbp]
  \centering
  \subfigure[]{\includegraphics[width=0.4\textwidth]{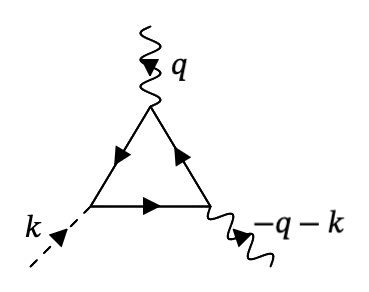}}
  \subfigure[]{\includegraphics[width=0.32\textwidth]{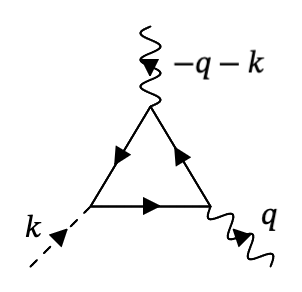}}
  \caption{subdiagram of diagram \ref{1c}}
  \label{sub of 1c}
\end{figure}

Next we calculate the diagram in Fig. (\ref{1c}).  This calculation can be done by first calculating the triangle diagram that yields the effective vertex $\phi FF $. Then inserting this for the triangle subdiagram in Figure~\ref{1c}. From the calculating the triangle diagram, we get the following effective coupling 
\begin{align}
  \mathrm{i}V_{\phi FF,\mathrm{EDM}}^{\mu\nu}\simeq\mathrm{i}\sum_{j}\frac{e^{2}}{24\pi^{2}}\frac{1}{m_{j}}\frac{v}{\Lambda}y_{\mathrm{S}}^{jj}\big[(k\cdot q -q^{2})\eta^{\mu\nu}-q^{\nu}k^{\mu}\big]~.
\end{align}
This allows us to replace $c_{\gamma}$ with $
\displaystyle\sum_{j}\frac{1}{96\pi^{2}}\frac{v}{m_{j}}y_{\mathrm{S}}^{jj} $ in $d_{2,\ell} $ in diagram~\ref{1c}. Due to the symmetry of the triangle subdiagram, a factor of $2 $ is included because of the possible exchanges of the photon lines within the subdiagram, see figure~\ref{sub of 1c}.  This then gives
\begin{align}
  d_{3,\ell}\simeq-2\times\sum_{j}\frac{e^{3}}{12\pi^{2}}\frac{v^{2}}{\Lambda^{2}}\frac{1}{m_{j}}y_{\mathrm{P}}^{ii}y_{S}^{jj}I_{2}.
\end{align}
A similar technique using subdiagrams to generate an effective coupling for $\mathrm{i}\phi \bar{\Psi}\gamma_{5}\Psi $ can be used in Figure~\ref{1d}, which gives  
\begin{align}
  \mathrm{i}\mathrm{V}_{\mathrm{i}\phi \bar{\Psi}\gamma_{5}\Psi,\mathrm{EDM}}\simeq\frac{3e^{4}}{4\pi^{2}}\frac{\tilde{c}_{\gamma}}{\Lambda}\slashed{k}\gamma_{5}\ln\bigg(\frac{\Lambda}{m_{i}}\bigg).
\end{align}
Again, EDM background contributions from the effective vertex $\phi F\tilde{F} $ are subdominant just as in $d_{2,\ell}, d_{3,\ell} $ and we get
\begin{align}
    \mathrm{i}M_{4,\ell\mathrm{EDM}}^{0}\simeq &8e^{3}\frac{1}{\Lambda^{2}}\int{\frac{\dd[4]{k}}{(2\pi)^4}}\;\tilde{\Delta}_{\phi}(k^{2})\frac{2m_{i}}{(k^{2}+q^{2})(2k\cdot\bar{p}+k^{2})}\bar{u}_{s_{2}}(p_{2})\notag\\ 
    &\qquad \times \frac{3e^{4}}{4\pi^{2}}\tilde{c}_{\gamma}c_{\gamma}(k^{0}k_{0}-k^{0}q^{2}/2m_{i})q_{\nu}\gamma^{0\nu}\gamma_{5}u_{s_{1}}(p_{1}),
\end{align}
with the following EDM, 
\begin{align}
    d_{4,\ell}\simeq-\frac{12e^{7}}{\pi^{2}}\frac{c_{\gamma}\tilde{c}_{\gamma}}{\Lambda^{2}}m_{i}I_{2}~.
\end{align}

Now we evaluate the integrals and collect the contributions to the lepton EDM. The occupation number density can be estimated using \eqref{occupation number density}. With this simplification, the integrals in \eqref{I_1} and \eqref{I_2} can be evaluated immediately giving the following leading order lepton EDM contribution\footnote{These contributions are independent of whether $q\ll k $ or $q\gg k $.}
\begin{align}
    \frac{d_{\ell}}{e}\simeq &\frac{\rho_{\phi}}{\Lambda^{2}m_{i}m_{\phi}^{2}}\Bigg[\frac{1}{2}y_{\mathrm{S}}^{ii}y_{\mathrm{P}}^{ii}\frac{v^{2}}{m_{i}^{2}}
    +8\pi\alpha y_{\mathrm{P}}^{ii}c_{\gamma}\frac{v}{m_{i}}+\sum_{j}\frac{\alpha}{3\pi}y_{\mathrm{P}}^{ii}y_{\mathrm{S}}^{jj}\frac{v^{2}}{m_{i}m_{j}}\notag\\ 
    &\qquad\qquad\quad
    +192\pi\alpha^{3}\tilde{c}_{\gamma}c_{\gamma}\ln\bigg(\frac{\Lambda}{m_{i}}\bigg)\Bigg],
\end{align}
where $\alpha=e^{2}/4\pi $ is the fine structure constant, and we have ignored higher order terms suppressed by $m_{\phi}/m_{i} $.

\section{Nucleon EDM}
\label{section 4}
To match the notation of \cite{CPV Axion review}, the ALP effective Lagrangian is modified from \eqref{general Lagrangian} so that it features a derivative ALP coupling to the quark axial current
\begin{align}
  \mathcal{L}_{\phi}^{\mathrm{QCD}}\supset &e^{2}\frac{\tilde{C}_{\gamma}}{\Lambda}\phi F\tilde{F}-\frac{v}{\Lambda}\frac{\partial_{\mu}\phi}{2m_{q}}\bar{q}\gamma^{\mu}y_{\mathrm{P}}\gamma_{5}q+g_{s}^{2}\frac{\tilde{C}_{g}}{\Lambda}\phi G\tilde{G}\notag\\
  &+e^{2}\frac{C_{\gamma}}{\Lambda}\phi FF+\frac{v}{\Lambda}\phi \bar{q}y_{\mathrm{S}}q+g_{s}^{2}\frac{C_{g}}{\Lambda}\phi GG,
  \label{general QCD Lagrangian}
\end{align}
where $q=\begin{pmatrix}
  u\\
  d
\end{pmatrix},\bar{q}=(\bar{u},\bar{d}) $ and $u,d $ are 4-component Dirac spinor of the up and down quark respectively, and $y_{\mathrm{P}}, y_{\mathrm{S}} $ are again hermitian matrices.

As was discussed in section~\ref{section 3}, chiral perturbative theory must be used below the GeV scale. The low energy effective Lagrangian containing ALP interactions with photons, nucleons and leptons is \cite{CPV Axion review,chiral Lagrangian} 
\begin{align}
    \mathcal{L}_{\phi}^{\chi\mathrm{pt}}\supset &e^{2}\frac{\tilde{c}_{\gamma}}{\Lambda}\phi F\tilde{F}+\mathrm{i}\tilde{C}_{\phi N}\frac{\partial_{\mu}\phi}{\Lambda} \bar{N}\gamma^{\mu}\gamma_{5}N+\mathrm{i}\frac{v}{\Lambda}y_{\mathrm{P}}^{ij}\phi \bar{\ell}_{i}\gamma_{5}\ell_{j}\notag\\
    &\quad+e^{2}\frac{c_{\gamma}}{\Lambda}\phi FF+\frac{C_{\phi NN}}{\Lambda} \phi\bar{N}N+\frac{v}{\Lambda}y_{\mathrm{S}}^{ij}\phi \bar{\ell}_{i}\ell_{j}
\end{align}
where $N=\begin{pmatrix}
  p\\ 
  n
\end{pmatrix}, \bar{N}=(\bar{p},\bar{n}) $ and $p,n $ are 4-component Dirac spinor of the proton and neutron respectively. 

The nucleon calculation results are very similar to the lepton calculation results. For the ALP-fermion-fermion interaction, we can merely replace the couplings, that is  $vy_{\mathrm{S}}^{ii}\rightarrow C_{\phi pp} $ and $vy_{\mathrm{P}}^{ii}\gamma_{5}\rightarrow \mp\tilde{C}_{\phi p}\slashed{k}\gamma_{5} $. However, in the nucleon sector,  the photon couplings to protons and neutrons are modified. The leading order\cite{photon-nucleon 1,photon-nucleon 2} interactions are
\begin{align*}
    \mathcal{L}_{\gamma N}\supset +e\bar{p}\gamma^{\mu}pA_{\mu}+\frac{\mathcal{C}_{n}}{4m_{n}}F_{\mu\nu}\bar{n}\sigma^{\mu\nu}n,
\end{align*}
where $\mathcal{C}_{n}\simeq 1.91 $ is measured low-energy constants.\cite{Review of particle physics}. 

\subsection{Proton EDM}
The leading contributions to the proton EDM also arises from the diagrams in Figure~\ref{EDM} with one CP-odd and one CP-even invariant vertex of $\phi $. 
However, the diagram in Figure~\ref{1c} is suppress by an additional proton mass unless the fermion loop is composed of lepton. Therefore, we ignore the case with a proton loop in Figure~\ref{1c}.

Diagram~\ref{1a} for the proton gives
\begin{align}
    \mathrm{i}M_{1,p}^{\mu}\simeq e\frac{1}{\Lambda^{2}}\int{\frac{\dd[d]{k}}{(2\pi)^d}}\;\tilde{\Delta}_{\phi}(k^{2})\frac{\bar{u}_{s_{2}}(p_{2})N^{\mu}_{1,p}u_{s_{1}}(p_{1})}{\big((\bar{p}+k)^{2}-m_{p}^{2}\big)^{2}}
\end{align}
where
\begin{align}
    N^{\mu}_{1,p}=C_{\phi pp}\tilde{C}_{\phi p}\big[(2k^{2}+8m_{p}^{2}+4k\cdot\bar{p})k^{\mu}+[\slashed{k},\gamma^{\mu}](k\cdot q)-4m_{p}\gamma^{\mu}(k\cdot q)\big]\gamma_{5}
\end{align}
Similar to the lepton case, only the term proportional to $k^{\mu}\gamma_{5} $ will contribute to the protons EDM. Again, the EDM arises from the 0-component of diagram~\ref{1a} and can be simplified to  
\begin{align}
  \mathrm{i}M_{1,p\mathrm{EDM}}^{0}\simeq & 2e\frac{C_{\phi pp}\tilde{C}_{\phi p}}{\Lambda^{2}}\int{\frac{\dd[4]{k}}{(2\pi)^4}}\;\tilde{\Delta}_{\phi}(k^{2})\frac{2m_{p}^{2}k^{0}+(k\cdot \bar{p})k^{0}}{m_{p}\big((\bar{p}+k)^{2}-m_{p}^{2}\big)^{2}}\times\bar{u}_{s_{2}}(p_{2})q_{\nu}\gamma^{0\nu}\gamma_{5}u_{s_{1}}(p_{1}),
\end{align}
which gives
\begin{align}
    d_{1,p}\simeq -2e\frac{C_{\phi pp}\tilde{C}_{\phi p}}{\Lambda^{2}}\big(2m_{p}I_{1}+I_{3}\big),
\end{align}
where $I_{1} $ is given by \eqref{I_1} and 
\begin{align}
  I_{3}=\int{\frac{\dd[4]{k}}{(2\pi)^4}}\;\tilde{\Delta}_{\phi}(k^{2})\frac{k_{0}k^{0}}{(k^{2}+2k\cdot\bar{p})^{2}}.
  \label{I_3}
\end{align}

The contribution to the EDM from Figure~\ref{1b} is 
\begin{align}
  \mathrm{i}M_{2,p\mathrm{EDM}}^{0}\simeq &\frac{8e^{3}}{\Lambda^{2}}\int{\frac{\dd[4]{k}}{(2\pi)^4}}\;\tilde{\Delta}_{\phi}(k^{2})\frac{1}{(k^{2}+q^{2})(2k\cdot\bar{p}+k^{2})}\bar{u}_{s_{2}}(p_{2})\notag\\ 
  &\qquad \times\big(c_{\gamma}\tilde{C}_{\phi p}2m_{p}(k^{0}k_{0}-k^{0}q^{2}/2m_{p})-2C_{\phi pp}\tilde{c}_{\gamma}\vb{k}^2\big)q_{\nu}\gamma^{0\nu}\gamma_{5}u_{s_{1}}(p_{1}),
\end{align}
giving an EDM of
\begin{align}
    d_{2,p}\simeq -16e^{3}\frac{c_{\gamma}\tilde{C}_{\phi p}}{\Lambda^{2}}m_{p}I_{2}~,
\end{align}
where $I_{2} $ is given by \eqref{I_2} and terms with $C_{\phi pp}\tilde{c}_{\gamma} $ are suppressed by the axion velocity, as we have mentioned in section~\ref{section 3}.

The proton EDM contribution of Figure~\ref{1c} is similar to the lepton EDM case. It can be found by replacing $c_{\gamma} $ with $\displaystyle \sum_{j}\frac{1}{96\pi^{2}}\frac{v}{m_{j}}y_{\mathrm{S}}^{jj} $ in $d_{2,p} $ and multiplying by a factor of two associated with the exchange of two photon line. This gives
\begin{align}
    d_{3,p}\simeq -2\times \sum_{j\in\{e,\mu,\tau\}}\frac{e^{3}}{6\pi^{2}}\frac{v}{m_{j}}\frac{y_{\mathrm{S}}^{jj}\tilde{C}_{\phi p}}{\Lambda^{2}}m_{p}I_{2}~.
\end{align}

Figure~\ref{1d} has the same form as the lepton case except the proton has a positive charge and a cut-off scale of $\Lambda_{\mathrm{ren}}\simeq m_{p} $. This leads to 
\begin{align}
  d_{4,p}\simeq -\frac{12e^{7}}{\pi^{2}}\frac{\tilde{c}_{\gamma}c_{\gamma}}{\Lambda^{2}}m_{p}\ln\bigg(\frac{\Lambda_{\mathrm{ren}}}{m_{p}}\bigg)I_{2}.
\end{align}

Using the occupation number density approximation found in \eqref{occupation number density} and integrating \eqref{I_1},\eqref{I_2} and \eqref{I_3}, we find the following contribution to the proton EDM
\begin{align}
    \frac{d_{p}}{e}\simeq&\frac{\rho_{\phi}}{\Lambda^{2}m_{p}m_{\phi}^{2}}
    \Bigg[\frac{1}{2}\tilde{C}_{\phi p}\frac{C_{\phi pp}}{m_{p}}
    +16\pi\alpha c_{\gamma}\tilde{C}_{\phi p}+\sum_{j\in\{e,\mu,\tau\}}\frac{2\alpha}{3\pi}y_{\mathrm{S}}^{jj}\tilde{C}_{\phi p}\frac{v^{2}}{m_{j}m_{p}}\notag\\ 
    &\qquad\qquad\quad
    +192\pi\alpha^{3}c_{\gamma}\tilde{c}_{\gamma}\ln \bigg(\frac{\Lambda_{\mathrm{ren}}}{m_{p}}\bigg)\Bigg].
\end{align}

\subsection{Neutron EDM}
The leading contribution to the neutron EDM are from the diagrams in Figure~\ref{1a},\ref{1b},\ref{1c}. The contribution from diagram~\ref{1d} is suppressed by multiple neutron-photon couplings, which to leading order is $\mathrm{i}\mathcal{C}_{n}q_{\nu}\gamma^{\mu\nu}/(2m_{n})$. This expression for the neutron photon interaction also highlights the fact that gauge invariance is automatically satisfied for the diagram in Figure~\ref{1a}. 

Diagram~\ref{1a} for the neutron gives
\begin{align}
    \mathrm{i}M_{1,n}^{\mu}\simeq-\frac{\mathcal{C}_{n}}{2m_{n}}\frac{C_{\phi nn}\tilde{C}_{\phi n}}{\Lambda^{2}}\int{\frac{\dd[d]{k}}{(2\pi)^d}}\;\tilde{\Delta}_{\phi}(k^{2})\frac{\bar{u}_{s_{2}}(p_{2})N^{\mu}_{1,n}u_{s_{1}}(p_{1})}{\big((\bar{p}+k)^{2}-m_{n}^{2}\big)^{2}},
\end{align}
with all the following terms giving rise to an EDM contribution,
\begin{align}
    N^{\mu}_{1,n\mathrm{EDM}}\simeq 8m_{n}\big[(k\cdot\bar{p}+k^{2})[\gamma^{\mu},\slashed{q}]-k^{\mu}[\slashed{k},\slashed{q}]-(4m_{n}^{2}-2k\cdot\bar{p}-k^{2})k^{\mu}\big]\gamma_{5}~.
\end{align}

Again, we take a gauge where $\vb{A} =0$.  This gives
\begin{align}
  \mathrm{i}M_{1,n}^{0}\simeq -4\mathcal{C}_{n}\frac{C_{\phi nn}\tilde{C}_{\phi n}}{\Lambda^{2}}\int{\frac{\dd[4]{k}}{(2\pi)^4}}\;\tilde{\Delta}_{\phi}(k^{2})\frac{k_{0}k^{0}}{(k^{2}+2k\cdot\bar{p})^{2}}\times \bar{u}_{s_{2}}(p_{2})q_{\nu}\gamma^{0\nu}\gamma_{5}u_{s_{1}}(p_{1}).
\end{align}
with a neutron EDM contribution of 
\begin{align}
    d_{1,n}\simeq 4\mathcal{C}_{n}\frac{C_{\phi nn}\tilde{C}_{\phi n}}{\Lambda^{2}}I_{3},
\end{align}
where $I_{3} $ is given by \eqref{I_3}.

The full expression for the matrix element from Figure~\ref{1b} is found in Appendix~\ref{appendix a}. The relevant part for the neutron EDM are shown below
\begin{align}
  \mathrm{i}M_{2,n\mathrm{EDM}}^{0}\simeq &-4e^{2}\frac{\mathcal{C}_{n}c_{\gamma}\tilde{C}_{\phi n}}{m_{n}\Lambda^{2}}\int{\frac{\dd[d]{k}}{(2\pi)^d}}\;\tilde{\Delta}_{\phi}(k)\frac{1}{2k\cdot\bar{p}+k^{2}}\bar{u}_{s_{2}}(p_{2})\notag\\
  &\quad\times\big[2m_{n}k^{0}-(2k\cdot\bar{p}+k^{2})\big]q_{\nu}\gamma^{0\nu}\gamma_{5}u_{s_{1}}(p_{1})
\end{align}
and results in an EDM of 
\begin{align}
  d_{2,n}\simeq &4e^{2}\frac{\mathcal{C}_{n}c_{\gamma}\tilde{C}_{\phi n}}{m_{n}\Lambda^{2}}\int{\frac{\dd[4]{k}}{(2\pi)^4}}\;\tilde{\Delta}_{\phi}(k^{2})\frac{k^{2}}{k^{2}+2k\cdot \bar{p}}\notag\\ 
  \simeq &4e^{2}c_{\gamma}\tilde{C}_{\phi n}\frac{\mathcal{C}_{n}m_{n}}{\Lambda^{2}}m_{\phi}^{2}\int{\frac{\dd[4]{k}}{(2\pi)^4}}\;\tilde{\Delta}_{\phi}(k^{2})\frac{1}{k^{2}+2k\cdot \bar{p}}~,
  \label{d2n}
\end{align}
Compared with $d_{2,\ell} $, the integral of $d_{2,n} $, again evaluated using \eqref{occupation number density}, is suppressed by $\order{m_{\phi}^{2}/m_{n}^{2}} $ and so can be ignored. We can also replace $c_{\gamma} $ with $\displaystyle \sum_{j}\frac{1}{96\pi^{2}}\frac{v}{m_{j}}y_{\mathrm{S}}^{jj} $ in $d_{2,n} $ to get $d_{3,n} $. Since this contribution is of the same order in $m_{\phi} $ as $d_{2,n} $, it can be ignored as well.

If we now evaluate the integral \eqref{I_3} using \eqref{occupation number density}, we have the following leading order contribution to the neutron EDM
\begin{align}
    \frac{d_{n}}{e}\simeq\frac{\rho_{\phi}}{\Lambda^{2}m_{n}m_{\phi}^{2}}
    \frac{\mathcal{C}_{n}}{e}\tilde{C}_{\phi n}\frac{C_{\phi nn}}{m_{n}}.
\end{align}

\section{Results}
\label{section 5}
We now apply our expressions for the EDMs from the previous section to determine the constraints on the ALP couplings.  Before proceeding, we need to further define our couplings using chiral perturbation theory. The coefficients of chiral perturbation theory are related to the microscopic parameters of \eqref{general QCD Lagrangian} via the following relations\cite{chiral Lagrangian}
\begin{align}
    &c_{\gamma}=C_{\gamma}+\cdots,\qquad \tilde{c}_{\gamma}=\tilde{C}_{\gamma}+\cdots\notag\\ 
    &\tilde{C}_{\phi p}=-\frac{v}{2m_{u}}y_{\mathrm{P}}^{uu}\Delta_{u}-\frac{v}{2m_{d}}y_{\mathrm{P}}^{dd}\Delta_{d}+\cdots,\notag\\ 
    &\tilde{C}_{\phi n}=-\frac{v}{2m_{u}}y_{\mathrm{P}}^{uu}\Delta_{d}+\frac{v}{2m_{d}}y_{\mathrm{P}}^{dd}\Delta_{u}+\cdots,\notag\\ 
    &C_{\phi pp}=\frac{v}{m_{u}}y_{\mathrm{S}}^{uu}\sigma_{u}+\frac{v}{m_{d}}y_{\mathrm{S}}^{dd}\sigma_{d}+\cdots,\notag\\ 
    &C_{\phi nn}=\frac{v}{m_{d}}y_{\mathrm{S}}^{uu}\sigma_{d}+\frac{v}{m_{u}}y_{\mathrm{S}}^{dd}\sigma_{d}+\cdots,\label{relation bewteen coefficient of chipt and microsopic parameter}
\end{align}
where ``$\cdots$'' represents terms related to gluon and the strange quark, which are not considered in this work. The values of $\Delta_{u},\Delta_{d},\sigma_{u},\sigma_{d} $ can be found in Ref\cite{Review of particle physics,QCD effect 1,QCD effect 2} and are 
\begin{align}
  &\Delta_{u}=0.86,\qquad\qquad\quad\;\;\Delta_{d}=-0.42,\notag\\
  &\sigma_{u}\simeq 10\sim 20 \;\mathrm{MeV},\qquad\sigma_{d}\simeq 20\sim 30 \;\mathrm{MeV}.
\end{align}

\subsection{Constraints From the Electron EDM}
Constraints on $d_{e} $ are obtained using the polar molecule ThO. The electron spin-precession frequency $\omega_{\mathrm{ThO}} $ is affected by both $d_{e} $ and CP-odd electron-nucleon interactions\cite{eEDM}
\begin{align}
    \omega_{\mathrm{ThO}}=1.2\; \mathrm{mrad}/\mathrm{s}\bigg(\frac{d_{e}}{10^{-29}e\;\mathrm{cm}}\bigg)+1.8\; \mathrm{mrad}/\mathrm{s}\bigg(\frac{C_{S}}{10^{-9}}\bigg),
\end{align}
with an experimental limit of $\omega_{\mathrm{ThO}}< 1.3\; \mathrm{mrad}/\mathrm{s} $ (90\% C.L.)\cite{EDM Experiment 1}. 
If we assume the entire dark matter density, $\rho_{\phi}\simeq 0.3\;\mathrm{GeV}/\mathrm{cm}^{3} $\cite{DM density}, is composed of an ALP, the experimental limit on the electron EDM will bound CP-violating invariant combinations of couplings. Our resulting constraints can be found in  Table~\ref{New bounds by electron EDM}. \footnote{We use eq.(42) of \cite{CPV Axion review} to evaluate the bounds of $y_{\mathrm{P}}^{ee}y_{\mathrm{S}}^{ee} $ at the scale $m_{\phi}=5\; \mathrm{GeV} $. It is unclear to us why our results differ by an order of magnitude from that found in \cite{CPV Axion review}.}

\begin{table}[htbp]
  \renewcommand{\arraystretch}{1.5}
  \centering
  \begin{tabular}{c|c|c|c}
      \hline 
      CPV invariant & New Bounds &Previous Bounds & Observable\\
      \hline 
      $y_{\mathrm{P}}^{ee}y_{\mathrm{S}}^{ee} $ & $2.7\times 10^{-23}\big(m_{\phi}/10^{-11} \mathrm{eV}\big)^{2} $ &$7.3\times 10^{-16} $ & $\omega_{\mathrm{ThO}}(d_{e}) $\\
      $y_{\mathrm{P}}^{ee}C_{\gamma} $ & $1.4\times 10^{-16} \big(m_{\phi}/10^{-11} \mathrm{eV}\big)^{2} $&$6.9\times 10^{-11} $ & $\omega_{\mathrm{ThO}}(d_{e}) $\\
      $y_{\mathrm{P}}^{ee}y_{\mathrm{S}}^{\mu\mu} $ & $ 1.4\times 10^{-17} \big(m_{\phi}/10^{-11} \mathrm{eV}\big)^{2} $& \diagbox{}{}& $\omega_{\mathrm{ThO}}(d_{e}) $\\ 
      $y_{\mathrm{P}}^{ee}y_{\mathrm{S}}^{\tau\tau} $ & $ 2.4\times 10^{-16} \big(m_{\phi}/10^{-11} \mathrm{eV}\big)^{2} $& \diagbox{}{}& $\omega_{\mathrm{ThO}}(d_{e}) $\\ 
      $C_{\gamma}\tilde{C}_{\gamma} $ & $ 5.2\times 10^{-8} \big(m_{\phi}/10^{-11} \mathrm{eV}\big)^{2} $&$6.2\times 10^{-3} $& $\omega_{\mathrm{ThO}}(d_{e}) $\\ \hline
  \end{tabular}
  \caption{A comparison of the new bounds presented here versus previous bounds found in \cite{CPV Axion review} for CP-violating invariants with $\Lambda=1\; \mathrm{TeV} $. The previous bounds are adjusted for $y_{\mathrm{P}}^{ee}y_{\mathrm{S}}^{ee} $ so they are consistent for $m_{\phi}\sim 10^{-11}\;\mathrm{eV} $. }
  \label{New bounds by electron EDM}
\end{table}

\subsection{Constraints From Nucleon EDM}
The current experimental bounds on $d_{p} $ and $d_{n} $ are $d_{p}<2.1\times 10^{-25}\;e\;\mathrm{cm} $ and $d_{n}<1.8\times 10^{-26}\;e\;\mathrm{cm} $ (90\% C.L.) \cite{EDM Experiment 2,EDM Experiment 3} respectively. They are used to bound CP-violating invariants on the nucleons. These bounds can be seen in Table~\ref{New bounds by nucleon EDM}.

\begin{table}[htbp]
  \renewcommand{\arraystretch}{1.3}
  \centering
  \begin{tabular}{|c|c|c|c|c|c|c|}
      \hline
          &$y_{\mathrm{S}}^{uu} $&$y_{\mathrm{S}}^{dd} $&$C_{\gamma} $&$y_{\mathrm{S}}^{ee} $&$y_{\mathrm{S}}^{\mu\mu} $& $y_{\mathrm{S}}^{\tau\tau} $ \\ \hline
      $y_{\mathrm{P}}^{uu} $&$1.8\times 10^{-13} $ &$1.4\times 10^{-13} $ &$7.8\times 10^{-10} $&$3.2\times 10^{-13}$&$6.5\times 10^{-11}$&$1.1\times 10^{-9}$ \\ \hline
      $y_{\mathrm{P}}^{dd} $& $1.9\times 10^{-13} $ &$1.5\times 10^{-13} $ &$3.4\times 10^{-9} $&$1.1\times 10^{-12}$ &$2.3\times 10^{-10}$&$3.8\times 10^{-9}$ \\ \hline
      $\tilde{C}_{\gamma} $&\diagbox{}{} &\diagbox{}{} & $3.0\times 10^{-1} $&\diagbox{}{}&\diagbox{}{}&\diagbox{}{}\\ \hline
  \end{tabular}
  \caption{New bounds on Jarlskog invariants obtained from the bounds on the neutron and proton EDMs assuming $\Lambda=1\;\mathrm{TeV} $ and $m_{\phi}=10^{-11}\;\mathrm{eV} $.}
  \label{New bounds by nucleon EDM}
\end{table}

As seen in Table~\ref{Bounds by nucleon EDM}, we find that the constraints on the Jarlskog invariants are stronger than recent results found in \cite{CPV Axion review} as long as $m_{\phi} $ is smaller than $10^{-11}\;\mathrm{eV} $. These constraints become even stronger as $m_{\phi} $ becomes smaller .

\begin{table}[htbp]
  \renewcommand{\arraystretch}{1.3}
  \centering
  \begin{tabular}{|c|c|c|c|c|c|c|}
      \hline
          &$y_{\mathrm{S}}^{uu} $&$y_{\mathrm{S}}^{dd} $&$C_{\gamma} $&$y_{\mathrm{S}}^{ee} $&$y_{\mathrm{S}}^{\mu\mu} $& $y_{\mathrm{S}}^{\tau\tau}$\\ \hline
      $y_{\mathrm{P}}^{uu} $&$2.8\times 10^{-12} $ &$1.8\times 10^{-12} $ &$8.0\times 10^{-8}$&$4.2\times 10^{-13}$&\diagbox{}{}&\diagbox{}{}\\ \hline
      $y_{\mathrm{P}}^{dd} $& $9.9\times 10^{-12} $ &$6.4\times 10^{-12}$ &$1.1\times 10^{-7}$&$4.2\times 10^{-13}$&\diagbox{}{}&\diagbox{}{}\\ \hline
      $\tilde{C}_{\gamma} $&$1.2\times 10^{-6} $ &$1.8\times 10^{-6} $ & $4.8\times 10^{-3} $&\diagbox{}{}&\diagbox{}{}&\diagbox{}{}\\ \hline
  \end{tabular}
  \caption{Bounds on Jarlskog invariants obtained from the bounds on the neutron and proton EDMs assuming $\Lambda=1\;\mathrm{TeV} $.\cite{CPV Axion review}}
  \label{Bounds by nucleon EDM}
\end{table}

\section{Conclusion}
\label{section 6}
In this work, we examined the effects of an ALP ultralight dark matter background on the EDM of the proton, the neutron, and leptons. For ultralight dark matter, the occupation number is
extremely high leading a Bose enhancement of the ALP propagation. This background effect generically leads to an enhancement to these process of order $\rho_{\phi}/(m_{\phi}^{2}m_{i}^{2}) $ and gives an enhancement of order $\sim 10^{4} $, if $m_{\phi}\sim 10^{-11}\;\mathrm{eV} $. This enhancement offsets the suppression coming from couplings and loop factors. Thus, if an ultralight ALP is dark matter, its couplings to the standard model particles will need additional suppression to avoid experimental detection. This technique has already been applied to the electrons $g-2$ contribution from ultralight dark matter, which resulted in new constraints on ALP dark matter couplings to the electron.  In this work, we examined the this effect on the electric dipole moment of the proton, neutron, and leptons. This resulted in new constraints on Jarlskog invariant combinations of CP-conserving and CP violation ALP couplings to standard model particles.  

Our calculation confirmed that the contribution to the proton, the neutron, and lepton EDMs from ALP CP conserving and CP violating couplings is enhanced by a factor of $\rho_{\phi}/(\Lambda^{2}m_{\phi}^{2}m_{i}) $ relative to vacuum contribution, for the EDM of particle $i$. 

Due to the enhancement from the background, we find new constraints on various Jarlskog invariants, if the ALPs mass is below $10^{-11}\,\mathrm{eV} $. Sixteen new constraints on Jarlskog invariants in all were found and are given in table~\ref{New bounds by electron EDM} and \ref{New bounds by nucleon EDM}. For example, $y_{\mathrm{P}}^{ee}y_{\mathrm{S}}^{ee}$ is 6 orders of magnitude stronger than the recent result\cite{CPV Axion review}, if $m_\phi$ is is dark matter and of order $10^{-11} $ eV.

\acknowledgments
We would like to thank Luca Di Luzio for useful discussion.

\newpage
\appendix

\section{Details of EDM Calculation}
\label{appendix a}
\subsection{The possible EDM term}
Since Lorentz invariance is broken due to the dark matter background, the effective form factor of the electromagnetic vertex for an EDM is not only $F_{3}(q^{3})q_{\nu}\gamma^{\mu\nu}\gamma_{5} $. Here we examine the non-relativistic limit of operators of the form $\bar{u}_{s_{2}}(p_{2})M u_{s_{1}}(p_{1}) $ with $M $ taken as $1,\gamma_{5},\gamma^{\mu},\gamma^{\mu}\gamma_{5},\gamma^{\mu\nu} $,
\begin{align}
    &\bar{u}_{s_{2}}(p_{2})1 u_{s_{1}}(p_{1})\simeq \xi_{s_{2}}^{\dag}\bigg[1-\frac{1}{4m^{2}}(\vb{p}_{1}\cdot\vb{p}_{2}+\mathrm{i}\boldsymbol{\sigma}\cdot(\vb{p}_{1}\times\vb{p}_{2}))\bigg]\xi_{s_{1}}\\ 
    &\bar{u}_{s_{2}}(p_{2})\gamma_{5} u_{s_{1}}(p_{1})\simeq -\xi_{s_{2}}^{\dag}\frac{\boldsymbol{\sigma}\cdot\vb{q}}{m}\xi_{s_{1}}\\ 
    &\bar{u}_{s_{2}}(p_{2})\gamma_{0} u_{s_{1}}(p_{1})\simeq \xi_{s_{2}}^{\dag}\bigg[1+\frac{1}{4m^{2}}(\vb{p}_{1}\cdot\vb{p}_{2}+\mathrm{i}\boldsymbol{\sigma}\cdot(\vb{p}_{1}\times\vb{p}_{2}))\bigg]\xi_{s_{1}}\\ 
    &\bar{u}_{s_{2}}(p_{2})\gamma^{0}\gamma_{5} u_{s_{1}}(p_{1})\simeq \xi_{s_{2}}^{\dag}\frac{\boldsymbol{\sigma}\cdot(\vb{p}_{1}+\vb{p}_{2})}{m}\xi_{s_{1}}\\ 
    &\bar{u}_{s_{2}}(p_{2})\boldsymbol{\gamma} u_{s_{1}}(p_{1})\simeq \xi_{s_{2}}^{\dag}\frac{(\vb{p}_{1}+\vb{p}_{2})+\boldsymbol{\sigma}\times\vb{q}}{2m}\xi_{s_{1}}\\ 
    &\bar{u}_{s_{2}}(p_{2})\boldsymbol{\gamma}\gamma_{5} u_{s_{1}}(p_{1})\simeq 2\xi_{s_{2}}^{\dag}\boldsymbol{\sigma}\xi_{s_{1}}\\ 
    &\frac{1}{2}\bar{u}_{s_{2}}(p_{2})[\gamma^{0},\boldsymbol{\gamma}] u_{s_{1}}(p_{1})\simeq \xi_{s_{2}}^{\dag}\frac{2\vb{q}+\boldsymbol{\sigma}\times(\vb{p}_{1}+\vb{p}_{2})}{2m}\xi_{s_{1}}\\ 
    &\frac{1}{2}\bar{u}_{s_{2}}(p_{2})[\gamma^{0},\boldsymbol{\gamma}]\gamma_{5} u_{s_{1}}(p_{1})\simeq 2\xi_{s_{2}}^{\dag}\boldsymbol{\sigma}\xi_{s_{1}}
\end{align}
where $q=p_{2}-p_{1} $ and 
\begin{align}
    u_{s}(p)=\begin{pmatrix}
      (1-p_{i}\sigma^{i}/2m)\xi_{s}\\ 
      (1+p_{i}\sigma^{i}/2m)\xi_{s}
    \end{pmatrix}.
\end{align}
So only when $M$ is a linear combination of $\gamma_{5}$, $[\gamma^{0},\vb{q}\cdot\boldsymbol{\gamma}]\gamma_{5}$ and $ \vb{q}\cdot\boldsymbol{\gamma}\gamma_{5} $ do we get $\bar{u}_{s_{2}}(p_{2})M u_{s_{1}}(p_{1})A_{0}(q) \propto \xi_{s_{2}}^{\dag}\vb{q}\cdot\boldsymbol{\sigma}\xi_{s_{1}}A_{0}(q)\simeq -\langle \vb{S}\rangle\cdot\vb{E} $, which gives a contribution to the EDM.

\subsection{Matrix elements from loop diagrams}
Here, we give the complete expressions for the matrix elements discussed in the draft.  

The entire matrix element for the diagram in Figure~\ref{1a} with the fermion being a lepton is 
\begin{align}
  \mathrm{i}M_{1,\ell}^{\mu}=&2e\frac{v^{2}}{\Lambda^{2}}\sum_{j}\int{\frac{\dd[4]{k}}{(2\pi)^4}}\;\tilde{\Delta}_{\phi}(k)\frac{1}{\big((\bar{p}+k)^{2}-m_{j}^{2}\big)^{2}}\times\bigg[2\Re(y_{\mathrm{S}}^{ij}y_{\mathrm{P}}^{ji})m_{j}k^{\mu}\bar{u}_{s_{2}}(p_{2})\gamma_{5}u_{s_{1}}(p_{1})\notag\\
  &\qquad-\mathrm{i}\Im(y_{\mathrm{S}}^{ij}y_{\mathrm{P}}^{ji})\bar{u}_{s_{2}}(p_{2})\Big[(k^{2}+m_{i}^{2}-m_{j}^{2})\gamma^{\mu}-2k^{\mu}\slashed{k}+m_{i}[\slashed{k},\gamma^{\mu}]\Big]\gamma_{5}u_{s_{1}}(p_{1})\bigg]
\end{align}

\begin{align}
  \mathrm{i}M_{1,\ell\mathrm{EDM}}^{\mu}\simeq 4e\frac{v^{2}}{\Lambda^{2}}\sum_{j}\Re(y_{\mathrm{S}}^{ij}y_{\mathrm{P}}^{ji})m_{j}\int{\frac{\dd[4]{k}}{(2\pi)^4}}\;\tilde{\Delta}_{\phi}(k^{2})\frac{\bar{u}_{s_{2}}(p_{2})k^{\mu}\gamma_{5}u_{s_{1}}(p_{1})}{\big((\bar{p}+k)^{2}-m_{j}^{2}\big)^{2}}.
  \label{lepton EDM 1}
\end{align}

For the diagram in Figure~\ref{1b}, again with leptons in the loop, is 
\begin{align}
  \mathrm{i}M_{2,\ell\mathrm{EDM}}^{\mu}\simeq&4e^{3}\frac{v}{\Lambda^{2}}\int{\frac{\dd[d]{k}}{(2\pi)^d}}\;\tilde{\Delta}_{\phi}(k)\frac{1}{(k^{2}+q^{2})(2k\cdot\bar{p}+k^{2})}\bar{u}_{s_{2}}(p_{2})\notag\\
  &\qquad\times\big(y_{\mathrm{P}}^{ii}c_{\gamma}k^{\mu}([\slashed{k},\slashed{q}]-2q^{2})+\mathrm{i}y_{\mathrm{S}}^{ii}\tilde{c}_{\gamma}\varepsilon^{\mu\nu\rho\sigma}q_{\rho}k_{\sigma}[\gamma_{\nu},\slashed{k}]\big)\gamma_{5}u_{s_{1}}(p_{1})
  \label{lepton EDM amplitude 2}
\end{align}

Next, we consider protons.  The diagram in Figure~\ref{1a} then gives 
\begin{align}
  \mathrm{i}M_{1,p}^{\mu}\simeq &eC_{\phi pp}\tilde{C}_{\phi p}\frac{1}{\Lambda^{2}}\int{\frac{\dd[d]{k}}{(2\pi)^d}}\;\frac{\tilde{\Delta}_{\phi}(k)}{\big((\bar{p}+k)^{2}-m_{p}^{2}\big)^{2}}\bar{u}_{s_{2}}(p_{2})\Big[(2k^{2}+8m_{p}^{2}+4k\cdot\bar{p})k^{\mu}\notag\\
  &\quad+[\gamma^{\mu},\slashed{k}](k\cdot q)-4m_{p}\gamma^{\mu}(k\cdot q)\Big]\gamma_{5}u_{s_{1}}(p_{1})
  \label{p 1}
\end{align}

The proton EDM contribution from the diagram in Figure~\ref{1b} is 
\begin{align}
  \mathrm{i}M_{2,p\mathrm{EDM}}^{\mu}\simeq 8e^{3}\frac{c_{\gamma}\tilde{C}_{\phi p}}{\Lambda^{2}}m_{p}\int{\frac{\dd[4]{k}}{(2\pi)^4}}\;\bar{u}_{s_{2}}(p_{2})\frac{k^{\mu}\big([\slashed{k},\slashed{q}]-2q^{2}\big)}{(k^{2}+q^{2})(2k\cdot\bar{p}+k^{2})}\gamma_{5}u_{s_{1}}(p_{1}),
\end{align}

The neutron EDM contribution is qualitatively different from the proton and lepton case due to the form of the nucleon-electromagnet vertex, $\mathrm{i}\mathcal{C}_{n}q_{\nu}\gamma^{\mu\nu}/(2m_{n})$. This leads to a neutron EDM contribution from the diagram in Figure~\ref{1a} of 
\begin{align}
  \mathrm{i}M_{1,n\mathrm{EDM}}^{\mu}\simeq &-4\mathcal{C}_{n}\frac{C_{\phi nn}\tilde{C}_{\phi n}}{\Lambda^{2}}\int{\frac{\dd[4]{k}}{(2\pi)^4}}\;\frac{\tilde{\Delta}_{\phi}(k^{2})}{(k^{2}+2k\cdot\bar{p})^{2}}\bar{u}_{s_{2}}(p_{2})\Big[(k\cdot\bar{p}+k^{2})[\gamma^{\mu},\slashed{q}]\notag\\ 
  &\qquad-k^{\mu}[\slashed{k},\slashed{q}]-(4m_{n}^{2}-2k\cdot\bar{p}-k^{2})k^{\mu}\Big]\gamma_{5}u_{s_{1}}(p_{1}),
\end{align}

For the diagram in Figure~\ref{1b}, we get is 
\begin{align}
  \mathrm{i}M_{2,n\mathrm{EDM}}^{\mu}\simeq &-8e^{2}\frac{\mathcal{C}_{n}c_{\gamma}\tilde{C}_{\phi n}}{2m_{n}\Lambda^{2}}\int{\frac{\dd[4]{k}}{(2\pi)^4}}\;\tilde{\Delta}_{\phi}(k^{2})\frac{1}{(k^{2}+q^{2})(k^{2}+2k\cdot \bar{p})}\bar{u}_{s_{2}}(p_{2})\notag\\ 
  &\qquad\big[4m_{n}^{2}k^{\mu}(q^{2}+k^{2})-(2k\cdot\bar{p}+k^{2})(q^{2}\eta^{\mu\rho}+k^{\mu}k^{\rho})q^{\sigma}\gamma_{\rho\sigma}\big]u_{s_{1}}(p_{1}).
\end{align}

Lastly, we comment on the form of the EDM used in the draft.  For $\bar{u}_{s_{2}}(p_{2})\gamma_{5}u_{s_{1}}(p_{1}) $, we can transform it to a more standard EDM form
\begin{align}
  \bar{u}_{s_{2}}(p_{2})\gamma_{5}u_{s_{1}}(p_{1})\simeq\frac{1}{2m}\bar{u}_{s_{2}}(p_{2})2\bar{p}^{0}\gamma_{5}u_{s_{1}}(p_{1})=\frac{1}{2m}\bar{u}_{s_{2}}(p_{2})q_{\nu}\gamma^{0\nu}\gamma_{5}u_{s_{1}}(p_{1}).
\end{align}

\subsection{Ward identities}
Gauge invariance is satisfied by the effective vertex for $\phi FF $ and $\phi F\tilde{F} $. So the diagrams in Figure \ref{1b}, \ref{1c} and \ref{1d} satisfy the Ward identities automatically. The gauge invariance of Figure~\ref{1a}, however, is not obvious. Here we consider if it satisfies the ward identities for both leptons and nucleons respectively.

\begin{figure}[htbp]
  \centering
  \subfigure[]{\includegraphics[width=0.4\textwidth]{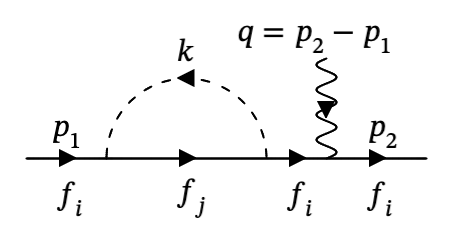}\label{2a}}
  \subfigure[]{\includegraphics[width=0.4\textwidth]{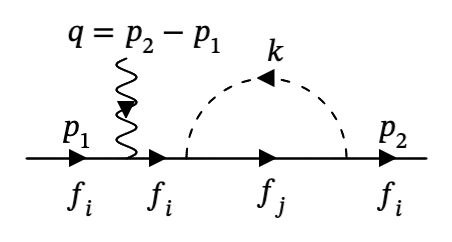}\label{2b}}
  \caption{Lepton Leg with CP-odd Loop}
  \label{CP-odd leg}
\end{figure}

Consider the diagrams with a lepton and a loop which is CP-odd (see figure~\ref{CP-odd leg}).  These diagrams have the same coupling as those found in figure~\ref{1a}. The resulting contribution of these diagrams is
\begin{align}
    \mathrm{i}M_{\ell,\text{leg}}^\mu=&2e\frac{v^{2}}{\Lambda^{2}}\sum_{j}\int{\frac{\dd[4]{k}}{(2\pi)^{4}}}\;\tilde{\Delta}_{\phi}(k^{2})\Bigg\{-\Re(y_{\mathrm{S}}^{ij}y_{\mathrm{P}}^{ji})\frac{m_{j}}{m_{i}}\frac{\bar{u}_{s_{2}}(p_{2})(q\cdot k)\gamma^{\mu}\gamma_{5}u_{s_{1}}(p_{1})}{\big[(\bar{p}+k)^{2}-m_{j}^{2}\big]^{2}}\notag\\
    &+\mathrm{i}\Im(y_{\mathrm{S}}^{ij}y_{\mathrm{P}}^{ji})\bar{u}_{s_{2}}(p_{2})\Bigg[\frac{\gamma^{\mu}\gamma_{5}}{(\bar{p}+k)^{2}-m_{j}^{2}}+\frac{\gamma^{\mu}(\slashed{p}_{1}+m_{i})\slashed{k}\gamma_{5}}{(p_{1}^{2}-m_{i}^{2})\big[(p_{1}+k)^{2}-m_{j}^{2}\big]}\notag\\
    &\qquad+\frac{\slashed{k}\gamma_{5}(\slashed{p}_{2}+m_{i})\gamma^{\mu}}{(p_{2}^{2}-m_{i}^{2})\big[(p_{2}+k)^{2}-m_{j}^{2}\big]}\Bigg]u_{s_{1}}(p_{1})\Bigg\}
\end{align}

If these contributions are added to the other diagrams calculated in the draft, we get
\begin{align*}
    q_{\mu}\big(\mathrm{i}M_{1,\ell}^{\mu}+\mathrm{i}M_{\ell,\text{leg}}^{\mu}\big)=0.
\end{align*}
and the ward identities are satisfied up to $\order{q^{3}}$. 

Next, consider the the diagrams in  Figure~\ref{CP-odd leg} for a proton. This then gives 
\begin{align}
  \mathrm{i}M_{p,\mathrm{leg}}^{\mu}\simeq eC_{\phi pp}\tilde{C}_{\phi p}\frac{v^{2}}{\Lambda^{2}}\int{\frac{\dd[4]{k}}{(2\pi)^{4}}}\;\frac{\tilde{\Delta}_{\phi}(k^{2})}{\big[(\bar{p}+k)^{2}-m_{p}^{2}\big]^{2}}\bar{u}_{s_{2}}(p_{2})\bigg(\gamma^{\mu}\gamma_{5}\frac{(k\cdot q)k^{2}}{m_{p}}\bigg)u_{s_{1}}(p_{1})
\end{align}
Again, we get 
\begin{align*}
    q_{\mu}\big(\mathrm{i}M_{1,\ell}^{\mu}+\mathrm{i}M_{\ell,\text{leg}}^{\mu}\big)=0.
\end{align*}
and the ward identities are satisfied up to $\order{q^{3}}$.

\section{Gauge Invariance with Background}
\label{appendix b}

According to the Dyson series, the S-matrix is
\begin{align*}
    S_{1,2,...;1',2',...}=\sum_{n=0}^{\infty}\frac{(-\mathrm{i})^{n}}{n!}\int\dd[4]x_{1}\cdots \dd[4]x_{n}(\Psi_{0},\cdots a_{2'}a_{1'}\times T\{\mathcal{H}(x_{1})\cdots\mathcal{H}(x_{n})\}a_{1}^{\dag}a_{2}^{\dag}\cdots \Psi_{0}),
\end{align*}
where $\Psi_{0} $ is ground state, $T $ is the time-ordering operator, and $\mathcal{H}(x) $ is the interaction Hamiltonian density. With a DM background, the ground state is no longer the vacuum but $|n(\vb{k})\rangle $, which is related to the dark matter particle distribution. The presences of this background leads to an additional term in the propagator of the dark matter particles\cite{Background UDM 1}. 

To show that the ward identities are generically satisfied in a background, if there are no in or out going dark matter particles, we present the following proof which closely follows Chapter 10 of Quantum Theory of Fields vol.1 of S. Weinberg.\cite{Weinber QFT 1}

\subsection*{Ward identities}
First, we define the electromagnetic vertex function $\Gamma^{\mu} $ of a charged particle by 
\begin{align}
    &\int\dd[4]x\dd[4]y\dd[4]z\;\mathrm{e}^{-\mathrm{i}k\cdot x}\mathrm{e}^{-\mathrm{i}p_{1}\cdot y}\mathrm{e}^{\mathrm{i}p_{2}z}(\Psi_{0},T\{J^{\mu}(x)\Psi_{n}(y)\bar{\Psi}_{m}(z)\}\Psi_{0})\notag\\ 
    \equiv &-\mathrm{i} q S'_{nn'}(p_{1})\Gamma_{n'm'}^{\mu}(p_{1},p_{2})S'_{m'm}(p_{2})\delta^{4}(k+p_{1}-p_{2}),
    \label{vertex function}
\end{align}
where
\begin{align}
    -\mathrm{i}S_{nm}'(p_{1})\delta^{4}(p_{1}-p_{2})\equiv \int\dd[4]y\dd[4]z\; \mathrm{e}^{-\mathrm{i}p_{1}\cdot y}\mathrm{e}^{\mathrm{i}p_{2}z}(\Psi_{0},T\{\Psi_{n}(y)\bar{\Psi}_{m}(z)\}\Psi_{0})
\end{align}
and $q $ is the electric charge of the particle.

In the limit of no interactions, these functions become
\begin{align*}
    S'(p)\rightarrow (\slashed{p}-m+\mathrm{i}\epsilon)^{-1},\qquad \Gamma^{\mu}(p_{1},p_{2})\rightarrow \gamma^{\mu}.
\end{align*}
The ward identities, which gives a relation between $\Gamma^{\mu} $ and $S' $, can be derived by the following identity
\begin{align}
  &\pdv{x^{\mu}}T\{J^{\mu}(x)\Psi_{n}(y)\bar{\Psi}_{m}(z)\}=T\{\partial_{\mu}J^{\mu}(x)\Psi_{n}(y)\bar{\Psi}_{m}(z)\}\notag\\
    &\qquad+\delta(x^{0}-y^{0})T\{[J^{0}(x),\Psi_{n}(y)]\bar{\Psi}_{m}(z)\}+\delta(x^{0}-z^{0})T\{\Psi_{n}(y)[J^{0}(x),\bar{\Psi}_{m}(z)]\}
\end{align}
Using 
\begin{align*}
    [J^{0}(\vb{x},t),\Psi_{n}(\vb{y},t)]=-q\Psi_{n}(\vb{y},t)\delta^{3}(\vb{x}-\vb{y}),\qquad [J^{0}(\vb{x},t),\bar{\Psi}_{m}(\vb{y},t)]=-q\bar{\Psi}_{m}(\vb{y},t)\delta^{3}(\vb{x}-\vb{y}),
\end{align*}
and charge conservation $\partial_{\mu}J^{\mu}=0 $, we can get
\begin{align}
  \pdv{x^{\mu}}T\{J^{\mu}(x)\Psi_{n}(y)\bar{\Psi}_{m}(z)\}=
    -q\delta^{4}(x-y)T\{\Psi_{n}(y)\bar{\Psi}_{m}(z)\}+q\delta^{4}(x-z)T\{\Psi_{n}(y)\bar{\Psi}_{m}(z)\}.
\end{align}
Inserting this in Fourier transform \eqref{vertex function} gives
\begin{align*}
    k_{\mu}S'(p_{1})\Gamma^{\mu}(p_{1},p_{2})S'(p_{2})\delta^{4}(k+p_{1}-p_{2})=\mathrm{i}S'(p_{2})\delta^{4}(k+p_{1}-p_{2})-\mathrm{i}S'(p_{1})\delta^{4}(k+p_{1}-p_{2})
\end{align*}
or
\begin{align}
  (p_{2}-p_{1})_{\mu}\Gamma^{\mu}(p_{1},p_{2})=\mathrm{i}S'^{-1}(p_{1})-\mathrm{i}S'^{-1}(p_{2}),
\end{align}
which is known as the Ward-Takahashi identity. 

\subsection*{Gauge invariance}
The quantities\footnote{The state $a $ and $b $ are the same as $\alpha $ and $\beta $, but with photons deleted.}
\begin{align}
    M^{\mu\mu'\cdots}_{\beta\alpha}(q,q',\cdots)\equiv \int\dd[4]x\int\dd[4]x'\cdots \mathrm{e}^{-\mathrm{i}q \cdot x}\mathrm{e}^{-\mathrm{i}q'\cdot x'}(\Psi_{b}^{-},T\{J^{\mu}(x)J^{\mu'}(x')\cdots\}\Psi_{a}^{+}), 
    \label{matrix of multi-J}
\end{align}
will vanish if contracted with any one of the photons four-momenta:
\begin{align}
    q_{\mu}M^{\mu\mu'\cdots}_{\beta\alpha}(q,q',\cdots)=q_{\mu'}'M^{\mu\mu'\cdots}_{\beta\alpha}(q,q',\cdots)=\cdots=0.
    \label{qM=0}
\end{align}
This statement is proved using conservation of electric charge.

By integration by parts, we have
\begin{align}
    q_{\mu}M^{\mu\mu'\cdots}_{\beta\alpha}(q,q',\cdots)=-\mathrm{i}\int\dd[4]x\int\dd[4]x'\cdots \mathrm{e}^{-\mathrm{i}q \cdot x}\mathrm{e}^{-\mathrm{i}q'\cdot x'}\Big(\Psi_{\beta}^{-},\pdv{x^{\mu}}T\{J^{\mu}(x)J^{\mu'}(x')\cdots\}\Psi_{\alpha}^{+}\Big),
    \label{qM} 
\end{align}

If there is only one current operator $J^{\mu} $ in \eqref{matrix of multi-J}, \eqref{qM} vanishes immediately by $\partial_{\mu}J^{\mu}=0 $. If there are more than two current operators, then the derivative in \eqref{qM} needs to be evaluated
\begin{align}
    \pdv{x^{\mu_{i}}}T\bigg\{\prod_{k}J^{\mu_{k}}(x_{k})\bigg\}=\sum_{j}\delta(x_{i}^{0}-x_{j}^{0})T\bigg\{[J^{0}(x_{i}),J^{\mu_{j}}(x_{j})]\prod_{k\neq j}J^{\mu_{k}}(x_{k})\bigg\}.
    \label{derivative of TJs}
\end{align}
And current operator $J^{\mu}(x) $ is an electrically neutral operator, that is $[J^{0}(\vb{x},t),J^{\mu}(\vb{y},t)]=0 $. Therefore, \eqref{derivative of TJs} vanishes, so \eqref{qM=0} is proved. Importantly, this makes no mention of the the vacuum state or the form of the propagators and so applies to the case with a background field as long as there are no external lines with background fields in the diagram.

\end{document}